\documentclass[twocolumn,aps,prl] {revtex4} 
\usepackage{graphicx}
\begin{document}
\pagestyle{empty} 
\title{Experimental Interpretation of Interfacial Separation and Squeezing Pressure }
\author{C. Yang$^1$,  B.N.J. Persson$^1$, J. Israelachvili$^2$ and K. Rosenberg$^2$}
\affiliation{$^1$Institut f\"ur Festk\"operForschung, Forschungszentrum J\"ulich, D-52425 J\"ulich, Germany}
\affiliation{$^2$Department of Chemical Engineering, University of California, Santa Barbara, California 93106, USA}

\begin{abstract}
The logarithmic relation between interfacial separation and squeezing pressure 
between randomly rough surfaces, has been predicted by both theory and 
experiment. However, the experimental 
slope between interfacial separation and logarithmic squeezing pressure, is
slightly bigger than that predicted by theory. Here we present a detailed 
explanation on the slope difference between theory and experiment. 

\end{abstract}
\maketitle


\begin{figure}
\includegraphics[width=0.45\textwidth,angle=0]{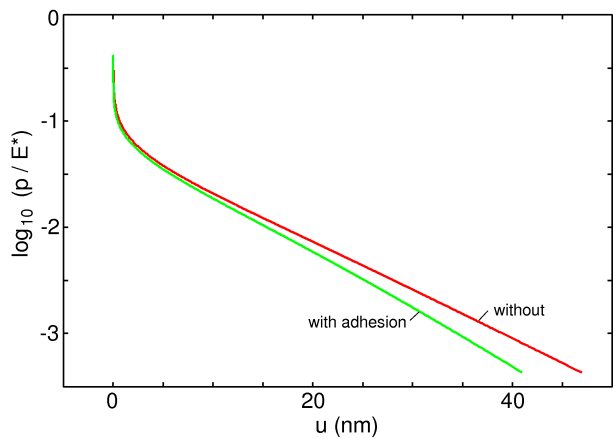}
\caption{\label{QQ}
Theoretical 
results for the logarithm
of the squeezing pressure $p$ (in units of $E^*=E/(1-\nu^2)$) as a function
of the interfacial separation $\bar u$. For the elastic modulus $E=0.9 \ {\rm GPa}$,
Poisson ratio $\nu = 0.5$ and
$\Delta \gamma = 0.1 \ {\rm J/m^2}$. Results are shown both with and without the adhesional
interaction. For the surface power spectrum (multiplied by a factor of 2) obtained from the
polymer surface topography
measured by M. Benz et al, see Ref.~\cite{MB} 
}
\end{figure}

Benz et al, see Ref.~\cite{MB}, have measured the (average) 
interfacial separation, $\bar u$, between two 
polymer surfaces with random roughness, as a function of the squeezing pressure $p$.
We can describe this system as an elastic solid (elastic modulus $E^* = E/(1-\nu^2),$ where
$E$ is half of the elastic modulus of the polymer) with a flat surface 
squeezed against a rigid surface with two times larger surface roughness
power spectrum than that of the original polymer surface.
In Fig.~\ref{QQ} we show the theoretical 
results for the logarithm
of the squeezing pressure $p$ (in units of $E^*=E/(1-\nu^2)$) as a function
of the interfacial separation $\bar u$. The elastic modulus $E=0.9 \ {\rm GPa}$,
Poisson ratio $\nu = 0.5$.
Results are shown both with adhesional interaction
(for the interfacial binding energy
$\Delta \gamma = 0.1 \ {\rm J/m^2}$),
and without the adhesional
interaction. The results without adhesion agree very well with the 
Finite Element Method (FEM) calculations of Pei et al\cite{Rob}.
Note that the absolute value of the 
slope of the curve with adhesion included is about $\sim 20\%$ higher than when the adhesion
is neglected. However, the experiment shows $\sim 50\%$ larger slope in spite of the fact that
the adhesional binding energy may be smaller than the one used in the calculation. 

\begin{figure}
\includegraphics[width=0.45\textwidth,angle=0]{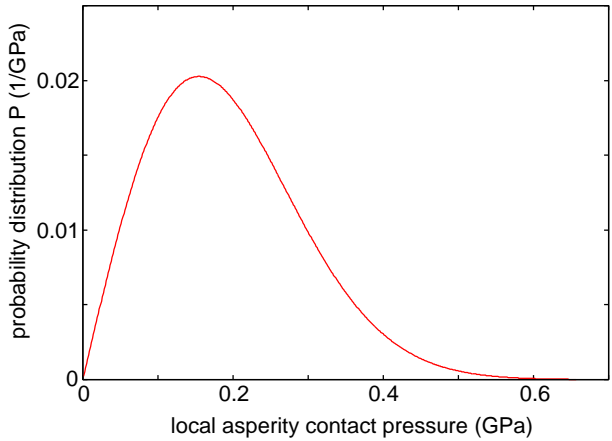}
\caption{\label{pressure.Probability}
The distribution $P(\sigma)$ of normal stress in the asperity contact areas. 
The nominal stress $p = 1 \ {\rm MPa}$.
The area under the
curve gives the relative contact area $A/A_0$.
For the elastic modulus $E=0.9 \ {\rm GPa}$ and the
Poisson ratio $\nu = 0.5$.
The calculation is for elastic contact without adhesion.
For the surface power spectrum (multiplied by a factor of 2) obtained from the
polymer surface topography
measured by M. Benz et al, see Ref.~\cite{MB} 
}
\end{figure}

The calculation above assumes purely elastic contact. However,
one may ask if the contact between the solids is elastic or if plastic yield will occur in 
some of the asperity
contact regions. To study this, in Fig.~\ref{pressure.Probability} we show
the distribution $P(\sigma)$ of normal stress in the asperity contact areas when 
the nominal stress $p = 1 \ {\rm MPa}$.
The area under the
curve gives the relative contact area $A/A_0$.
The calculation is for elastic contact without adhesion. If the yield stress 
(or rather the indentation hardness)
of the solids is larger than $\sim 0.5 \ {\rm GPa}$, one expects negligible plastic flow.
The (macroscopic) indentation hardness (Vickers test) of typical polymers such as PMMA, POM and PE are
$\sim 0.1-0.3 \ {\rm GPa}$, so for these polymers one would expect important plastic yield to
occur. This would effectively smoothen the surfaces and result in a steeper curve in
Fig. \ref{QQ}, in agreement with observations. However, the indentation hardness of the polymer used 
in the experiment in Ref.~\cite{MB} is not known to us.

In summary, the absolute slope the 
theory predicts is smaller than that observed in the experiment by Benz et al.\cite{MB}.
It might be due to that, in the experiment, 
plastic yield occurs in some asperity contact regions which effectively smoothen the surface.
However, there are also a number of uncertainties or corrections to the analysis of Ref.~\cite{MB}, which we
now discuss.

There are three sources of corrections to be made to the published results\cite{MB}.
Two of them decrease (the absolute value of) the
slope of the ${\rm log} p - \bar u$ relation, and one increases the slope. 
We can refer to them as (i) the refractive index correction 
(also third body or amount of confined material correction), 
(ii) the deformation correction (the flattening or increase in the radius of the macroscopic surface, 
not the asperities), and 
(iii) the elasticity correction (other contributions to the elasticity 
of the system than the force-measuring spring). 
Some of these effects depend on the way the forces, distances and 
surface geometry are measured or sensed, i.e., on the experimental 
setup and method. It might be of interest to know exactly 
what these are in our case because they could arise in other systems 
that involve the experimental measurement or control of the adhesion and 
friction of rough surfaces, including the compaction of rough powders, 
and maybe even Geckos! In order of importance in our SFA experiments, this is 
what we have estimated (had we measured more parameters or recorded more 
images at the time of the experiments we could have done accurate quantitative 
analysis of the results, but we didn't think them to be significant, 
although we now see that they are). So here goes:

(i) When calculating the distances we need to input the refractive index of the medium of 
the film between the two confining reference surfaces. Now for rough surfaces this is not 
a simple matter because there is no sharply defined reference plane, e.g., like a 
solid-liquid interface. So the film or gap which was made up of asperities and air or 
liquid was assumed to have a constant refractive index equal to that of the bulk polymer 
material, i.e., of the asperities. At small gaps (high compression), this should be true, 
but at large gaps, where the forces are first detected, the refractive index would be lower 
than the one assumed in the calculation by a factor that depends on the volume fraction 
occupied by the asperities. Assuming that at the tail end of the forces $80 \%$ 
of the gap is filled with polymer while at the smallest distances 
$100 \%$ is occupied we conclude that our smallest distances were 
calculated correctly but the larger ones were underestimated by up to $10 \%$ which results 
in a slope that is up to $35 \%$ smaller than what we plotted. 
This correction would increase the decay length which, interestingly, brings the 
results closer to the theoretical results. The extent of 
this correction depends on the absolute range of the forces, and  
the shorter the range the smaller the percentage correction. 

(ii) The flattening correction at maximum compression (effectively larger $R$), 
which was also ignored in the analysis, would cause an additional 
reduction in the slope by about the same amount, i.e., $\sim 20 \%$.

(iii) The stiffness correction (i.e., including the Hertzian contribution of the 
deforming glue layers to the stiffness of the normal force-measuring spring, 
where the Hertzian stiffness is not constant but increases with the load) would 
increase the slope, but this appears to be a small effect, of order $\sim 5 \%$.

\end{document}